\documentclass{aa}
\usepackage{graphicx}
\usepackage{natbib,multirow}
\bibpunct{(}{)}{;}{a}{}{,}
\usepackage{txfonts}
\begin{document}
   \title{SS\,433\,: a phenomenon imitating a Wolf-Rayet star\thanks{Based 
        on observations with ISO, an ESA project with instruments funded by ESA Member States (especially the PI countries: France, Germany, the Netherlands, and the United Kingdom) and with the participation of ISAS and NASA.}}


   \author{Y. Fuchs
          \inst{1,}\inst{2}
          \and
          L. Koch Miramond\inst{1}
        \and
        P. \'Abrah\'am\inst{3}
          }

   \offprints{Y. Fuchs \\ \email{yfuchs @ cea.fr}}

   \institute{Service d'Astrophysique, CEA/Saclay, 
        Orme des Merisiers B\^at. 709, 91191 Gif-sur-Yvette, France
         \and
        Universit\'e Joseph Fourier, LAOG, BP 53, 38041 Grenoble 
                cedex 9, France
         \and
             Konkoly Observatory, P.O. box 67, 1525 Budapest, Hungary
             }

   \date{Received 12 October 2004 / Accepted 3 August 2005}

   \abstract{We present mid-infrared (2--12\,$\mu$m) spectra 
   of the microquasar SS\,433 obtained
   with the Infrared Space Observatory (spectroscopic mode of ISOPHOT
   and ISOCAM).
   We compare them to the spectra of
   four Wolf-Rayet stars: WR\,78, WR\,134, WR\,136, and WR\,147 in the
   same wavelength range. The mid-infrared spectrum of SS\,433 
   mainly shows \ion{H}{i} and \ion{He}{i} emission lines and is very
   similar to the spectrum of WR\,147, a WN8(h)+B0.5V binary. 
        The 2--12\,$\mu$m continuum emission of SS\,433
   corresponds to optically thin and partially optically thick
   free-free emission, from which we calculate a mass loss rate of
   $2-3\times 10^{-4} M_\odot \,\mathrm{yr}^{-1}$
   if the wind is
   homogeneous and a third of these values if it is clumped. This is
   consistent with a strong stellar wind from a WN star. However,
   following recent studies concluding that the mass donor star of
   SS\,433 is not a Wolf-Rayet star, we propose that this strong wind
   out flows from a geometrically thick envelope of material that
   surrounds the compact object like a stellar atmosphere, imitating the
   Wolf-Rayet phenomenon. This wind could also wrap the mass donor
   star, and at larger distances ($\sim40$\,AU), 
   it might form a dust envelope
   from which the thermal emission, detected with ISOPHOT at 25\,$\mu$m 
    and 60\,$\mu$m, would originate.
   This wind also probably feeds the material that is
	ejected in the orbital plane of
   the binary system and that forms the equatorial outflow detected in
   radio at distances $>100$\,AU.

   }

   \maketitle
%

\section{Introduction}

        SS\,433 was the first microquasar (an X-ray binary with
        relativistic jets) discovered in the 1970's, when that
        designation had not even been invented yet. It was first
        discovered as a star with strong H$\alpha$ emission lines and so
        was included in the Stephenson and Sanduleak catalog
        \citep{stephsand77} as object number 433. It was later
        associated to a variable X-ray and radio source
        \citep{zwitter89}.  Detailed study of its visible spectrum
        revealed 
        that this source is unique. 
	For a complete review of SS\,433,
        we refer the reader to \citet{margon84}, \citet{zwitter89},
        \citet{verm96}, and to the introduction of
        \citet{giesmcswain02}.

        There are two sets of optical emission lines in the spectrum
        of SS\,433: the first set corresponds to 
        the so-called ``stationary'' lines \citep{margon79}
        showing normal
        Doppler shift movements, including the strong H$\alpha$ lines, with
        a period of 13~days.
        The second set regroups the ``moving''
        lines, those that are less intense (by about 1/3) but 
	that show huge Doppler
        shifts ($+50\,000\,$Km.s$^{-1}$ and $-35\,000\,$Km.s$^{-1}$ at
        the maximum elongation) corresponding to relativistic
        velocities with a period of about 162~days \citep{margon79}.
        The ``kinematic model'' is generally the one chosen to explain these
        unusual observations: SS\,433 is an X-ray binary system
        orbiting in 13~days and emitting relativistic jets where the
        moving lines are formed. 
	These jets undergo a precession
        movement in 162 days. The parameters of this system were
        recently re-calculated by \citet{eiken01}, who found:
        $P_\mathrm{orb}=13.08$\,d,
        $P_\mathrm{prec}=162.375\pm0.011$\,d, velocity of the
        ejections $v=0.2647\pm0.0008$\,c, inclination of the jet axis
        to the line of sight $i=78.05\degr\pm0.5\degr$, and opening
        angle of the precession cone $\theta=20.93\degr\pm0.08\degr$.
        The ejections and their precession movement are indeed
        observed in the radio images; see e.g. \citet{hjellming81} and
        \citet{verm93}. The source has always been observed ejecting
        material, meaning that this is not only 
	the first microquasar discovered but
        also the only one with continuous ejections, and the only object
        known to show evidence of \emph{ions} accelerated to relativistic
        velocities (0.26\,c).

        Despite intensive studies, the nature of the
        two stars in this binary system remains 
        uncertain.
        No emission line  has been identified as clearly belonging to
        the donor star of this X-ray binary, some of them being
        suspected to be formed close to the compact object.
        Numerous mass ratios were estimated leading to a low-mass or
        high-mass X-ray binary with either a neutron star or a black
        hole candidate.
        The presence of Wolf-Rayet-like lines 
        added to the very luminous
        continuum in the visible and near-IR ranges led several people 
	to associate
        the donor star to a Wolf-Rayet or Of star
        \citep{murdin80,vandenheuvel80,hut81}.
        However, the coordinated optical and X-ray light curves of SS\,433 show
        that the X-ray eclipse corresponds to the minimum visual magnitude
        \citep{stewart87}; thus a thick accretion disc is inferred, 
        which
        is more luminous in the visible range than the donor star.
        Only recently, 
        \citet{gieshuang02} 
        detected absorption features in
        the blue spectrum  suggesting that the donor star is
        an A-type evolved star. In the more detailed spectroscopic
        study of \citet{hillwig04}, 
        the donor star is consistent with an
        \mbox{A3-7\,I} type with $M=10.9\pm3.1\,M_\odot$,
        and the compact object 
        is a low-mass black hole candidate
        with $M_\mathrm{X}=2.9\pm0.7\,M_\odot$.

        The distance of SS\,433 
        has also remained uncertain for a long time. 
        From complete observations
        of the radio emitting materials ejected in the relativistic
        jets 
        and after taking  the kinematic model into account, 
        \citet{blundell04} eventually determined 
        this distance as $d=5.5\pm0.2$\,kpc.

        As the origin of the spectrum (lines and continuum) 
        of SS\,433 in the visible and
        near-infrared ranges appears 
        quite complicated for understanding clearly, 
         the idea was to look for characteristic lines in
        the mid-IR range which could help to understand 
        the nature and the origin of the emissions better and possibly 
        to constrain 
        the nature of the donor star further.
        In this article we present observations of SS\,433 with the
        Infrared Space Observatory (ISO) in both spectroscopic and
        photometric modes in the 2--12\,$\mu$m range at different
        epochs and at 25\,$\mu$m \& 60\,$\mu$m. We compare the emission
        line spectra to the spectra of four Wolf-Rayet star of WN
        type (Sect.~3). We also study the continuum emission of
        SS\,433 in Sect.~4 
        and calculate the mass loss and radius of
        the corresponding free-free emission. In Sect.~5 we discuss
        the constraints on the nature of the mass donor star provided
        by our mid-IR observations, we describe the
        phenomenon imitating
        a Wolf-Rayet star, and we consider the possible large scales
        behaviour of the strong mass outflow.


\section{Observations and data reduction}


     \begin{table*}[!t]
     \caption{Log of the observations of SS\,433 with ISO. The $\Psi$ represents the precessional phase from \citet{goranskii98}, $\Phi_\mathrm{p}$ and $\varphi$ are, respectively, the precessional and orbital phase from \citet{kemp86}. The flux densities were given by the PHOT data reduction, were directly measured from the CAM images, or were roughly estimated from the ISOPHOT spectra (continuum level). See the text for the corresponding uncertainties.}
     \label{tabobsSS433}
     \begin{minipage}{\hsize}
     \begin{tabular}{lrcccllll}
\hline
\hline
TDT & civil date and in Julian Day
  & $\Psi$ & $\Phi_\mathrm{p}$ & $\varphi$
                & mode & $\lambda$ range (FWHM) & aperture & flux density\\
 & & & & & & and filters & or PFOV & range (mJy)\\
\hline
35300404 & 3 Nov. 1996  \ 2\,450\,391.195  & 0.41 & 0.84 & 0.11 & PHT40\footnote{spectrophotometry with ISOPHOT} 
        & 2.47--4.87 $\mu$m  ($\sim 0.04\,\mu m$) & $24'' \times 24''$ 
                & 380--350\\
\hspace{0.5cm}'' & '' \hspace{0.4cm} '' \hspace{0.9cm} '' \hspace{0.4cm} '' \hspace{0.5cm} & '' & '' & '' & PHT40
        & 5.84--11.62 $\mu$m  ($\sim 0.1\,\mu m$) & $24'' \times 24''$ 
                & 330--180\\
51200601 & 11 Apr. 1997 \ 2\,450\,550.272  & 0.39 & 0.81 & 0.27 & PHT40 
        & 2.47--4.87 $\mu$m  ($\sim 0.04\,\mu m$) & $24'' \times 24''$ 
                & 710--590\\
\hspace{0.5cm}'' & '' \hspace{0.4cm} '' \hspace{0.9cm} '' \hspace{0.4cm} '' \hspace{0.5cm} & '' & '' & '' & PHT40
        & 5.84--11.62 $\mu$m  ($\sim 0.1\,\mu m$) & $24'' \times 24''$ 
                & 530--280\\
51200602 & '' \hspace{0.4cm} '' \hspace{0.5cm} \ 2\,450\,550.298  & '' & '' & '' & PHT03\footnote{multi-filter photometry  with ISOPHOT} 
        & P1 12: $11,89 \pm 3.255$\,$\mu$m  & $18''$ 
                & 354\\
\hspace{0.5cm}'' & '' \hspace{0.4cm} '' \hspace{0.9cm} '' \hspace{0.4cm} '' \hspace{0.5cm} & '' & '' & '' & PHT03
        & P2 25: $23.81 \pm 4.59$\,$\mu$m  & $52''$ 
                & 432\\
51200603 & '' \hspace{0.4cm} '' \hspace{0.5cm} \ 2\,450\,550.303  & '' & '' & '' & PHT22\footnote{multi-filter photometry with the far-IR camera of ISOPHOT}
        & C100 60: $60.8 \pm 11.95$\,$\mu$m & $43'' \times 43''$\footnote{per pixel, giving a total field of view of $130'' \times 130''$}
                & 390\\
51801005 & 17 Apr. 1997 \ 2\,450\,556.016  & 0.43 & 0.85 & 0.71 & PHT40
        & 2.47--4.87 $\mu$m  ($\sim 0.04\,\mu m$) & $24'' \times 24''$ 
                & 350--450\\
\hspace{0.5cm}'' & '' \hspace{0.4cm} '' \hspace{0.9cm} '' \hspace{0.4cm} '' \hspace{0.5cm} & '' & '' & '' & PHT40
        & 5.84--11.62 $\mu$m  ($\sim 0.1\,\mu m$) & $24'' \times 24''$ 
                & 320--200\\
52402506 & 23 Apr. 1997 \ 2\,450\,561.939  & 0.47 & 0.89 & 0.16 & PHT40
        & 2.47--4.87 $\mu$m  ($\sim 0.04\,\mu m$) & $24'' \times 24''$ 
                & 230--330\\
\hspace{0.5cm}'' & '' \hspace{0.4cm} '' \hspace{0.9cm} '' \hspace{0.4cm} '' \hspace{0.5cm} & '' & '' & '' & PHT40
        & 5.84--11.62 $\mu$m  ($\sim 0.1\,\mu m$) & $24'' \times 24''$ 
                & 260--180\\
70900204 & 24 Oct. 1997 \ 2\,450\,746.268  & 0.60 & 0.02 & 0.25 & CAM01\footnote{raster imaging with ISOCAM}
        & LW3:   12--18 $\mu$m & $3'' \times 3''$\footnote{per pixel}
                & 124\\
\hspace{0.5cm}'' & '' \hspace{0.4cm} '' \hspace{0.9cm} '' \hspace{0.4cm} '' \hspace{0.5cm} & '' & '' & '' & \hspace{0.4cm}'' & LW7:  8.5--10.7 $\mu$m & '' \hspace{0.4cm} '' \          & 148\\
\hspace{0.5cm}'' & '' \hspace{0.4cm} '' \hspace{0.9cm} '' \hspace{0.4cm} '' \hspace{0.5cm} & '' & '' & '' & \hspace{0.4cm}'' & LW9:  14--16 $\mu$m & '' \hspace{0.4cm} '' \                     & 123\\
\hspace{0.5cm}'' & '' \hspace{0.4cm} '' \hspace{0.9cm} '' \hspace{0.4cm} '' \hspace{0.5cm} & '' & '' & '' & \hspace{0.4cm}'' & LW6:  7--8.5 $\mu$m & '' \hspace{0.4cm} '' \                     & 207\\
\hspace{0.5cm}'' & '' \hspace{0.4cm} '' \hspace{0.9cm} '' \hspace{0.4cm} '' \hspace{0.5cm} & '' & '' & '' & \hspace{0.4cm}'' & LW8:  10.7--12 $\mu$m & '' \hspace{0.4cm} '' \           & 139\\
\hspace{0.5cm}'' & '' \hspace{0.4cm} '' \hspace{0.9cm} '' \hspace{0.4cm} '' \hspace{0.5cm} & '' & '' & '' & \hspace{0.4cm}'' & LW4:  5.5--6.5 $\mu$m & '' \hspace{0.4cm} '' \           & 226\\
71100514 & 26 Oct. 1997 \ 2\,450\,748.294  & 0.61 & 0.03 & 0.41 & CAM01
        & LW2:  5--8.5 $\mu$m & $1.5'' \times 1.5''$\,$^f$
                & 423\\
\hline
        \end{tabular}
       \end{minipage}
        \end{table*}

        We searched for observations of SS\,433,        
        $R.A. = 19^\mathrm{h}11^\mathrm{m}49^\mathrm{s}.57$    
        $Dec. = +04\degr58'57''.8$ (J2000),
        in the archives of the
        Infrared Space Observatory\footnote{http://www.iso.vilspa.esa.es/ida/} 
(ISO, \citealt{kessler96}) and
        found several of them in 1996 and 1997 as shown in
        Table~\ref{tabobsSS433}. The main observations were achieved
        with the spectroscopic mode of the photometer on board ISO,
        ISOPHOT \citep{lemke96} at four different epochs. The other
        observations were carried out with photometric modes of
        ISOPHOT and with another instrument of ISO: the
        infrared camera ISOCAM \citep{cesarsky96}.

        We looked in the archives of the Green Bank Interferometer
        (GBI) and at the quick-look results provided by the All-Sky
        Monitor / Rossi X-ray Timing Explorer (ASM/RXTE) team, in
        order to know the state of SS\,433 in radio and X-rays,
        respectively, at the time of the observations with ISO. The GBI
        2.25\,GHz and 8.3\,GHz lightcurves show a normal activity in
        the radio, i.e. only small oscillations with no flare during
        the ISO observations. The X-ray level 
	detected by ASM is low (a few counts/s), as usual, and has no 
	flaring event.

	In the visible range the spectrum of SS\,433 shows
        strong H$\alpha$ emission lines and looks like the spectrum of a
        Wolf-Rayet star, so
        we searched in the ISO archives for
        observations of WR-type stars, particularly those of late WN-type
        which are less evolved and so still emit hydrogen lines. We found
        four observations of well-defined WN stars (see
        table~\ref{tabWR}) with ISOSWS \citep{degraauw96} the short
        wavelength spectrometer of ISO, which has a far better spectral
        resolution than ISOPHOT.

           \subsection{ISOPHOT wide-band filters}

SS\,433 was observed at 12, 25, and 60~$\mu$m with ISOPHOT 
on 11 April 1997 (see Table~\ref{tabobsSS433}). 
The measurements at 12 and 25~$\mu$m were
performed in triangular chopped mode; i.e. the line-of-sight was
switched periodically between SS433 and two background positions
placed symmetrically at $\pm 120''$ from the source.  At 60~$\mu$m,
also measured in chopped mode, only
one background position was selected at 180$''$ from SS433.

The data reduction was performed using the ISOPHOT Interactive
Analysis Software Package V9.1 (PIA, \citealt{gabriel97}). After
corrections for non-linearities of the integration ramps, an 8-point
signal pattern was created from each observation by overplotting and
averaging the basic blocks of the observation (the repeated
background+source cycles); for details see \citet{abraham01}. The
signals of the patterns were transformed to a standard reset interval
of 1/4 s, and an orbital dependent dark current was subtracted. The
signals were corrected for non-linearities of the detectors by
applying the signal linearization corrections, as available in March
2001.  From the patterns a [source--background] difference signal was
then extracted.  In most cases the measured difference signal
underestimates the real signal due to short term detector transients:
therefore, a correction was applied for signal loss. This correction is
a function of the chopping frequency and the measured difference
signal. The flux calibration of the 12 and 60~$\mu$m measurements was
performed by comparison with the on-board fine calibration source
(FCS), which was also measured in chopped mode. At 25~$\mu$m the
detector's actual sensitivity could be reliably predicted from the
orbital position of the observation, and an orbital dependent default
responsivity was applied. Finally the derived flux densities were
corrected for the finite size of the aperture by using the standard
correction values.  Colour corrections were applied for each
measurement.

The flux uncertainty at 25~$\mu$m was estimated to be better than 30\%. 
At 12~$\mu$m the selected aperture of $18''$ was smaller than the
standard one in which the ISOPHOT calibration programme was
executed. At 60~$\mu$m the actual responsivity of the detector can
deviate significantly from an average default figure, and its value
had to be derived from the accompanying FCS measurement, thereby introducing
an additional uncertainty factor. Due to these issues, the accuracies
of the 12 and 60~$\mu$m measurements are expected to be more in the
30-50\% range.

           \subsection{ISOPHOT spectral mode}
Low resolution mid-infrared spectra of SS433 were obtained 
on 3 November 1996, 11, 17, and 23 April 1997, 
(see Table~\ref{tabobsSS433}) with the ISOPHOT-S
sub-instrument. The spectra covered the 2.5--4.9~$\mu$m and 5.9--11.7~$\mu$m
wavelength ranges with 
spectral resolutions of about 0.04\,$\mu$m and 0.1\,$\mu$m 
and sensitivities of 26\,mJy and 10\,mJy, respectively.
The observations were performed in triangular chopped mode with two
background positions located at $\pm$90$''$. The orientation of the
chopping direction on the sky depends on the date of observation and is
different for each spectrum. The dwelling time per chopper position was
128\,s, and each measurement consisted of 4 OFF1--ON--OFF2--ON cycles and
lasted 2048\,s.

The ISOPHOT-S data were reduced in three steps:\\
(1) We used the Phot Interactive Analysis (PIA, \citealt{gabriel97})
software V9.1 to filter out cosmic glitches in the raw data and to
determine signals (in V/s) by performing linear fits to the
integration ramps.  After a second de-glitching step, performed on the
signals, a dark current value appropriate to the orbital position of
the individual signal was subtracted.  Then we averaged all
non-discarded signals to derive a signal per chopper step.\\
(2) An [ON--OFF] signal averaged for the whole measurement was
determined by applying a 1-dimensional Fast Fourier Transformation
algorithm; for the application of FFT methods for ISOPHOT data
reduction see \citet{haas00}).\\
(3) The [ON--OFF] difference signals were calibrated by applying a
signal dependent spectral response function dedicated to chopped PHT-S
observations \citep{laureijs}.

In order to verify our data reduction scheme (which is not completely
standard due to the application of the FFT algorithm) and
to estimate the level of the calibration uncertainties, we reduced HD\,184400,
an ISOPHOT standard star observed in a similar way as SS433.
The results are very consistent with the model prediction for the star,
and we estimate that the systematic
uncertainty of our calibration is not greater than 10$\%$.
    However, as PHOT-S was designed for broader spectral features than
    the lines observed in this study, the measured line flux might be
    underestimated because the ISOPHOT-S pixels are separated by small
    gaps; according to Table 4.5 of the ISOPHOT Handbook
    \footnote{http://www.iso.vilspa.esa.es/manuals/HANDBOOK/pht\_hb/node3.html}
    about 75\% of the line flux is measured if the line is centred on
    a pixel's wavelength; and about 43\% if the line is centred between
    two pixels' wavelengths.

    \begin{table*}[!t]
    \caption{Main properties of the WR stars to be compared with
        SS\,433} 
    \label{tabWR}
    \begin{minipage}{\hsize}
    \begin{tabular}{lccccl}
\hline
\hline
Name & WR type\footnote{from \citet{smithlf96}} & binary & distance & $A_V$\footnote{from \citet{vanderhucht01} except for WR\,147} & references\footnote{vdH01 = \citet{vanderhucht01} ; 
C95a = \citet{crowther95II} ; M99 = \citet{morel99} ; CS96 = \citet{crowthersmith96} ; C95b = \citet{crowther95IV} ; SH99 = \citet{stevens99} ; Mor00 = \citet{morris00}}\\ %
\hline
WR\,78 = HD\,151932 & WN7h WNL & no & 2.0 kpc & 1.48--1.87 & vdH01 ; C95a\\ 
WR\,134 = HD\,191765 & WN6b WNE-s & possible & 1.74 / 2.1\,kpc & 1.22--1.99 & vdH01 / M99 ; CS96\\
WR\,136 = HD\,192163 & WN6b(h) WNE-s & possible & 1.26 / 1.8\,kpc & 1.35--2.25 & vdH01 / SH99 ; C95b\\ 
WR\,147 & WN8(h) WNL & B0.5\,V at 0.55$''$ & $630\pm70$\,pc & 11.6 / 11.2 & vdH01 / Mor00\\ 
\hline
     \end{tabular}
     \end{minipage}
     \end{table*}

           \subsection{ISOCAM wide-band filters}

        SS\,433 was observed with the camera of ISO, with several
        large band filters on 24 and 26 October 1997 (see
        Table~\ref{tabobsSS433}).
        The ISOCAM data were reduced with the Cam Interactive Analysis
        software (CIA, \citealt{Delanay00}) version 4.0, following the
        standard processing outlined in \citet{starck99}.  First a
        dark correction was applied, then a de-glitching to remove
        cosmic ray hits, followed by a transient correction
        to take memory effects into account. Pixels showing remnants of
        these effects were masked, as well as insufficiently
        lit side pixels. The jitter correction was applied for the
        $1.5''\times 1.5''$ resolution image. The flat-field
        correction used the automatic evaluation.  Then
        individual images were combined into the final raster map, 
        and finally the pixel values were converted into milli-Jansky
        flux densities. No colour correction was applied.
        We estimated to $\sim20$\% the total uncertainty of the ISOCAM
        observed flux densities.

           \subsection{ISOSWS data on WR stars}

        We took the SWS spectra of the four WR stars from the archive;
        i.e. they were processed by the SWS pipeline (officially
        Off-Line Processing = OLP) version 9. In order to compare them
        to the PHOT-S spectra of SS\,433, we had to degrade their
        resolution and keep only the wavelength range of the PHOT-S
        modes. The convolution to the PHOT-S spectral resolution was
        performed with IDL routines available on the VILSPA home
        page\footnote{http://www.iso.vilspa.esa.es/instr/SWS/ ``SWS vs
        PHT-S Convolution Tool" written by B. Schulz}.



\section{The emission lines: comparison with Wolf-Rayet stars}

\begin{figure}[!t]
   \includegraphics[width=\linewidth]{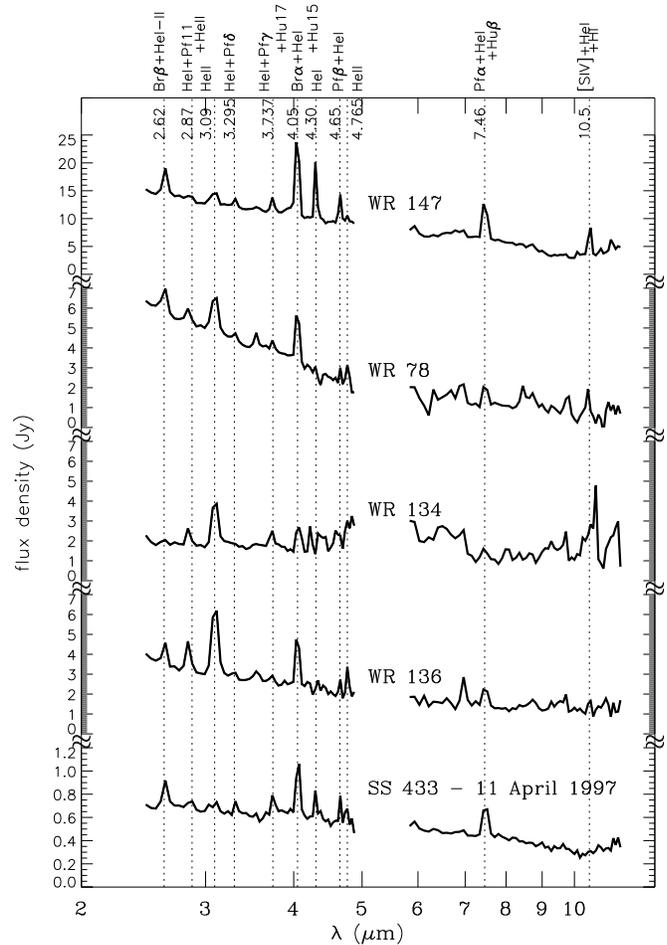}
   \caption{Observed (i.e. absorption not corrected) spectra of
   SS\,433 with ISOPHOT-S on 11 April 1997 and of four Wolf-Rayet
   stars with ISOSWS rebinned to the PHOT-S resolution and wavelength
   range. The $\lambda$ axis has a logarithmic scale, in order to
   better separate the different lines in the low $\lambda$ part of
   the spectra.}
    \label{figSS433WR}
    \end{figure}

        As we summarized in the introduction, the nature of 
	the visible and near-IR
        spectrum of SS\,433 is difficult to interpret, 
	and till
        recently no line could be used to determine the nature of the
        donor star in this X-ray binary. ISO 
        provided
         unique spectra of this
        object in the mid-IR range.
	Since SS\,433 has for a long time been
        strongly suspected to be a Wolf-Rayet star of
        WN type 
        and its optical spectrum indeed resembles those of WN stars,
        we decided to compare its
        emission line spectra to that of several WN stars
        found in the ISO archives and rebinned to the same
        spectral resolution.
        The main properties of these four WR are shown in
        Table~\ref{tabWR}, and their observed spectra are displayed in
        Fig.~\ref{figSS433WR}, along with the spectrum of SS\,433 as
        observed on 11 April 1997, since it has the strongest emission
        among the available spectra of this source.

        Figure~\ref{figSS433WR} shows the observed spectra i.e. with no
        correction for absorption by the interstellar medium.  All the
        emission lines were identified using \citet{morris00} and only
        the lines present in at least three spectra of SS\,433 and at
        2.5\,$\sigma$ above the continuum were taken into account. No
        absorption line was found. The emission lines are mainly
        \ion{H}{i} lines blended with He\,\textsc{i} or
        He\,\textsc{ii} lines. There is also a weak He\,\textsc{i}
        line at 4.30\,$\mu$m and two weak He\,\textsc{ii} lines at
        3.09\,$\mu$m and 4.765\,$\mu$m, but no metallic line is
        detected.  It is clear in Fig.~\ref{figSS433WR} that the
        spectrum of SS\,433 is closest to the one of WR\,147, 
        a WN8(h) star, 
        although the isolated \ion{He}{i} (4.30\,$\mu$m) and
        \ion{He}{ii} (3.09\,$\mu$m) lines are more marked in this star
        than in SS\,433; conversely, the other WR stars differ by a
        strong He\,\textsc{ii} 3.09\,$\mu$m line.

   \begin{figure}[!t]
   \includegraphics[width=\linewidth]{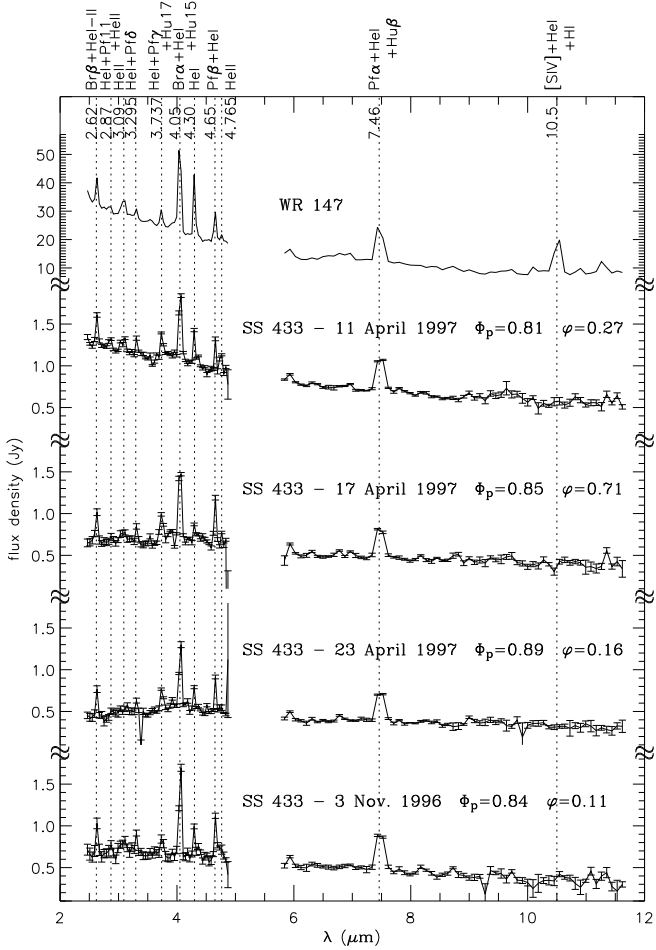}
   \caption{Dereddened (i.e. absorption corrected) spectra of WR\,147
   and SS\,433, using the law of \citet{lutz96} and $A_V=11.2$ \& $A_V=8$,
   respectively. $\Phi_\mathrm{p}$ and $\varphi$ are the precessional
   and orbital phases from \citet{kemp86}, as described in the text
   (Sect.~\ref{sectvar}). In the spectra of SS\,433, some absurd points
   due to instrumental defaults are enhanced by the dereddening, as at
   the upper edge of the 2--5\,$\mu$m range, and thus they have very
   large error bars going off of the graphics.}
    \label{figSS433WR147}
    \end{figure}

        The similarities between the spectrum of WR\,147 and the four
        spectra of SS\,433 also appear clearly once the spectra are
        dereddened, as in Fig.~\ref{figSS433WR147} where we corrected
        the interstellar extinction
        using the law from \citet{lutz96} (see the next section concerning the
        choice of this law) and a visual absorption of $A_V=8$ for
        SS\,433 \citep{margon84} 
        and of $A_V=11.2$ for WR\,147 \citep{morris00}.
%
        The spectra of SS\,433 show the same
        general shape as the one of WR\,147 with the same lines and
        comparable relative intensities between the latters, except
        for the 10.5\,$\mu$m line, which is absent in SS\,433. 
        \citet{smithhouck01} studied 8--13\,$\mu$m spectra of several
        WN8 and WN9 stars, where this  [\ion{S}{iv}] 10.5\,$\mu$m 
	line can be either very weak
        or not detectable and may be drowned in the noise of the spectrum.
        However, the spectra of SS\,433 are not very noisy around
        10.5\,$\mu$m, and the sensitivity of the detector
        ($\sim$\,10\,mJy) should enable us to detect this line if
        present. 
	 The absence of this line is 
        surprising, if we are dealing with a strongly ionized, dense
        outflow in SS433, which has a common physical mechanism to WN
        stars.  Indeed, one would expect the presence of mid-IR fine
        structure lines, such as [\ion{S}{iv}], unless the ionization
        in the relevant region (around a density of $10^5$ to 
	$10^6$~cm$^{-3}$) differs from that of WR stars. Given the similar
        ionization in the He line spectrum, one might suspect that the
        ionization of the plasma is comparable.
        Thus in the mid-IR, SS\,433 (only) looks like a late WN star (WNL)
        of WN8 
        type, which is relatively
        H-rich (H/He\,$\sim0.5-2$, \citealt{crowther02}).
%
        Note that WN8 stars are known to be the most active subclass of
        Wolf-Rayet stars because of their high levels of spectroscopic
        and photometric variability (see \citealt{morris00} and
        references therein). Note also that WR\,147 is a Wolf-Rayet
        binary (WN8+B0.5V) system with strong colliding wind that has
        been observed in radio \citep{williams97}.




        The measurement of line widths and position
        are limited by the spectral resolution
        of the spectrocopic mode of ISOPHOT i.e. FWHM $\sim 0.04\ \&\ 
        0.1 \,\mu$m for the 2--5\,$\mu$m and 6--12\,$\mu$m ranges, 
        respectively. Thus it is not possible to detect any Doppler
        shift due to the orbital movement. Concerning
        the 0.26\,c relativistic motion in the jets, there is no 
        evidence of such a Doppler shift in our spectra when comparing 
        the different precesional phases observed, 
        so there is no indication of infrared emission from the jets.
%
	Because of the limited spectral resolution and the blended lines,
	we could neither study the properties of the lines (equivalent
        width, shape, etc...)
	nor measure the elemental abundances.




\section{The continuum}

   \subsection{Dereddening}
\begin{figure}[!tp]
    \includegraphics[width=\linewidth]{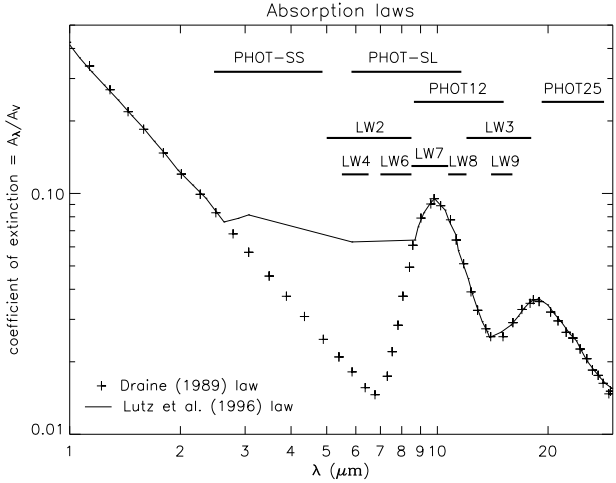}
    \caption{Comparison of the absorption law by
        \cite{draine89} with the \citet{lutz96} law in the mid-IR. The
        coefficient of extinction is defined by $A_\lambda / A_V$,
        where $A_V$ is the extinction in the optical V (0.55\,$\mu$m)
        band and $A_\lambda$ the extinction at wavelength
        $\lambda$. The wavelength ranges of the ISOPHOT and ISOCAM
        filters used in this article are overplotted.}  
   \label{figabs}
   \end{figure}

        In order to study the continuum emission of SS\,433, we have to
        deredden the observed values, i.e. to correct them from the
        absorption along the line of sight toward the source.  What we
        call dereddened flux $F_\mathrm{der}$ is linked to the
        observed flux $F_\mathrm{obs}$ by the formula:
        \mbox{$F_\mathrm{der}\,=\,F_\mathrm{obs}\,\times\,10^{+0.4A_{\lambda
        }}$}, where $A_{\lambda}$ is the extinction at wavelength
        $\lambda$, calculated as $A_V$ times the extinction
        coefficient at $\lambda$ given by an extinction law 
	(see Fig.~\ref{figabs}).  As
        SS\,433, with $l = 39\degr41'38.8"$ and $b = -02\degr14'40.6"$, is
        located in the galactic plane at a distance of 
        5.5\,kpc, this
        absorption is not negligible. The problem is that the visible
        absorption $A_V$ is not constrained very well for SS\,433.

        By fitting the optical continuum slope from UV to near-IR with
        the assumption of various thermal models (blackbody underlying
        spectral types or/and free-free emission) or using infrared
        spectrophotometry, many authors found $A_V\sim8$ (see
        references in \citealt{murdin80} and \citealt{margon84}).
        Evaluations using recent X-ray data are not restricting enough, as
        demonstrated in \citet{marshall02} on their model fitting of
        the Chandra grating X-ray spectra of SS\,433.  If they use a simple
        power-law model (a poor fit to the data), then they get
        $N_\mathrm{H} = 9.5 \times 10^{21}$\,cm$^{-2}$, which implies
        $A_V=5.3$ using the \citet{predschm95} formula:
        $N_\mathrm{H}(\mathrm{cm}^{-2}) / A_V = 1.79 \times
        10^{21}$. When they use the X-ray line ratios and
        simultaneously fit $N_\mathrm{H}$, they get $2.2\times
        10^{22}$\,cm$^{-2}$, which implies $A_V=12.29$; hence, the
        range of values one will find in the literature.  Also from
        Chandra observations, \citet{namiki03} just fitted a power-law
        + gaussian lines and found $N_\mathrm{H} = (1.0 - 1.3)
        \times 10^{22}$\,cm$^{-2}$, which gives $A_V = 6.1-7.3$\,mag.
%
        In the following we will only show spectra dereddened with
        $A_V=8$.
        Note that using lower values such as $A_V=5.3$ doesn't change
         our results significantly (see Sect.~\ref{sectmdot}).


        The extinction of the interstellar medium is highly wavelength
        dependent, even in the mid-IR range, but there are different
        absorption laws available depending on the galactic line of sight. 
        The law described by
        \citet{draine89} is commonly used, except for regions toward
        the Galactic Centre for which \citet{lutz96} showed that a
        flatter law (between 2.5 and 8.5\,$\mu$m) has to be
        used. Figure~\ref{figabs} shows the difference between these two
        laws and the wavelength range of our ISO observations. The Lutz law is
        also used in some particular cases like some dense
        H\,\textsc{ii} regions or some objects with a carbon-rich
        environment.
        As an example, \citet{moneti01}  used the Lutz law to
        deredden their 2.5--8\,$\mu$m ISO data on the Quintuplet
        Cluster, although this young cluster is in the vicinity of the
        Galactic Centre.
%
%
        \citet{indebetouw04} presented very recent
        extinction measurements in the 1.25--8.0\,$\mu$m range along
        the $l=42\degr$ and $l=284\degr$ line of sight in the galactic
        plane, which are similar and consistent with measurements of
        \citet{lutz96} toward the Galactic Centre. This shows that the
        Lutz law could almost be universal or, at least, more common
        than previously thought.
        \citet{dwek04} attributes this law to the presence of
        relatively short metallic needles in the interstellar medium
        that could form in the supernova ejecta and in quiescent
        O-rich stellar outflow. SS\,433 may have undergone a supernova
        explosion to form its compact object, as it is surrounded by
        the W50 radio nebula \citep{dubner98}, which is possibly a supernova
        remnant. Thus 
	the Lutz law is likely to be valid in the case of SS\,433,  
	so we used it in this study.
         Moreover when we dereddened the spectra with the Draine law, the
        continuum showed a complicated shape with a narrow ``bump'' in
        the \mbox{7--10\,$\mu$m} range, which we were unable to fit correctly,
        even when using a lower $A_V$ value.


   \subsection{Variability}
        \label{sectvar}

\begin{figure*}[!t]
   \includegraphics[width=9cm]{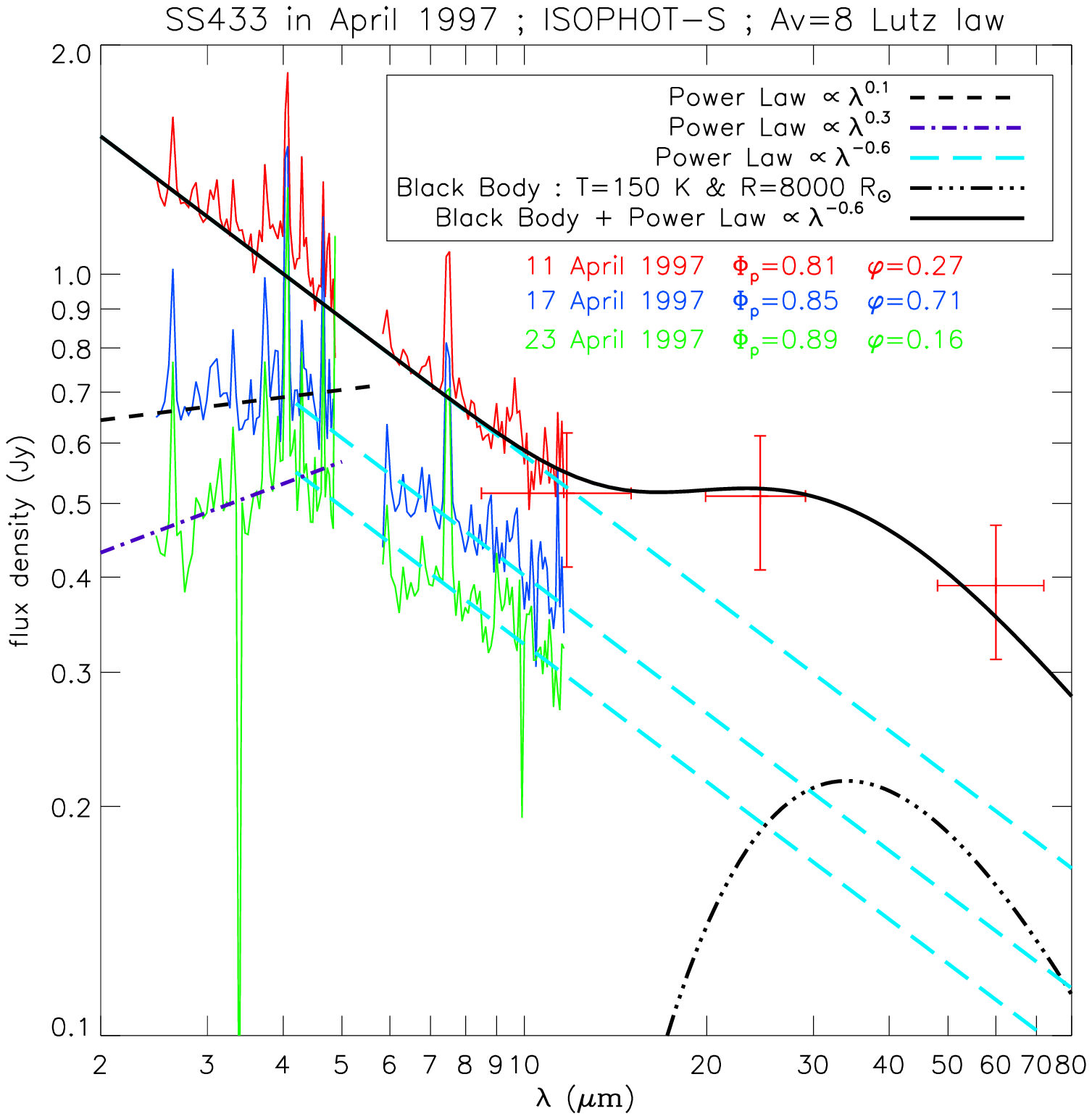}
        \includegraphics[width=9cm]{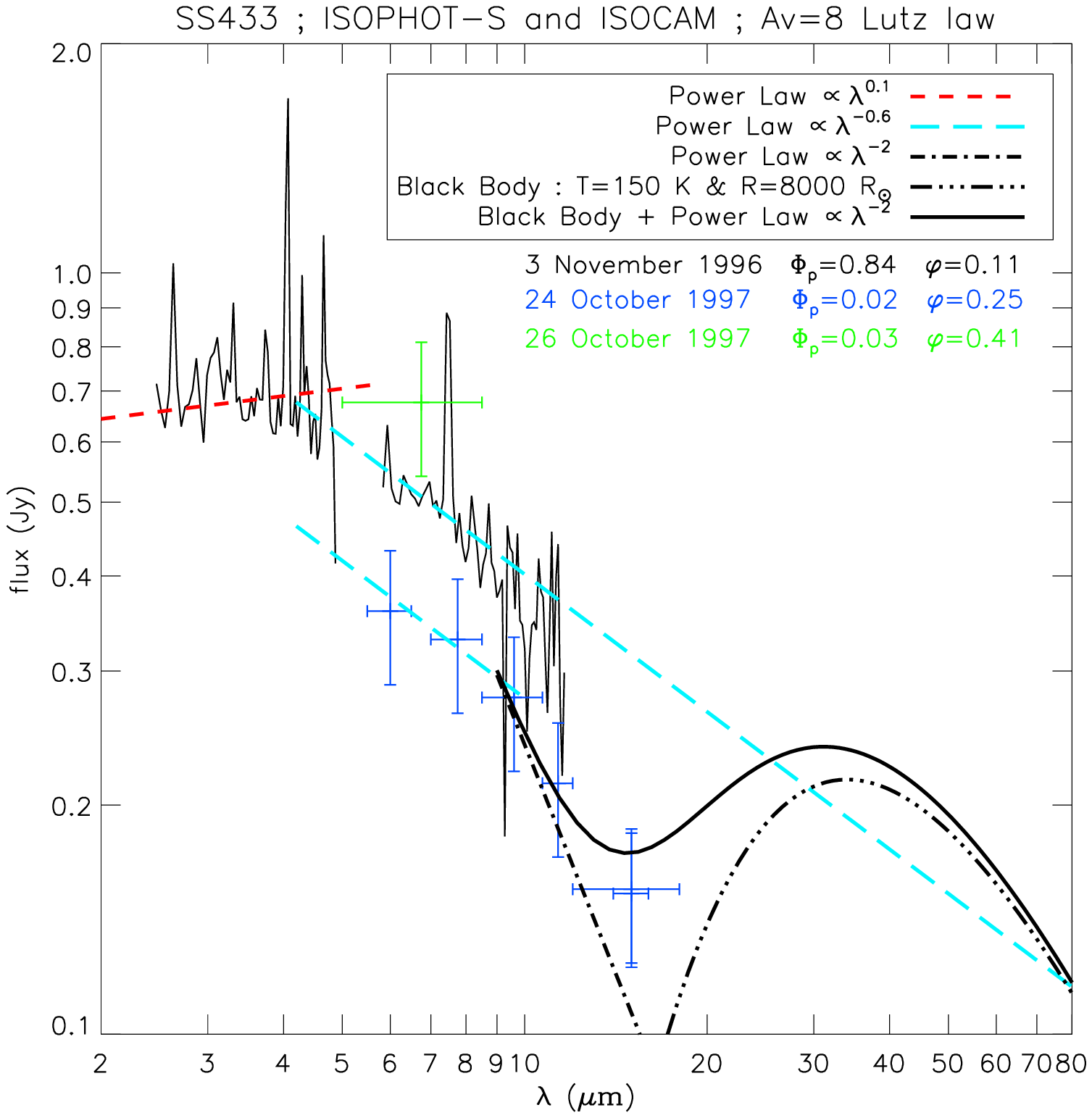}
   \caption{Fit of the continuum emission of the dereddened spectra of 
	SS\,433 with power laws proportional to $\lambda^{0.1}$ 
	(short dashes) or $\lambda^{0.3}$ (dash dot, left side) 
	or $\lambda^{-2}$ (dash dot, right side) or/and $\lambda^{-0.6}$ 
	(long dashes) and a black body emission with $T=500$\,K, 
	$R=8000\,R_\odot$ at a distance of 5.5\,kpc (dash triple dot).
	In the left figure the solid line fits the emission of 11 April 1997  
	(black body + power law $\propto \lambda^{-0.6}$), and in the right 
	figure it fits the emission of 24 October 1997 for 
	$\lambda > 9\,\mu$m (black body + power law $\propto \lambda^{-2}$).}
   \label{figcont}
   \end{figure*}

        Figure~\ref{figcont} shows all the ISO observations of SS\,433 once
        dereddened with $A_V=8$ and using the law of
        \citet{lutz96}. The continuum level varies from one
        observation to another, and we can try to explain these
        variations using the orbital and precessional phases.

\begin{figure}[!t]
   \centering
   \includegraphics[width=6.5cm]{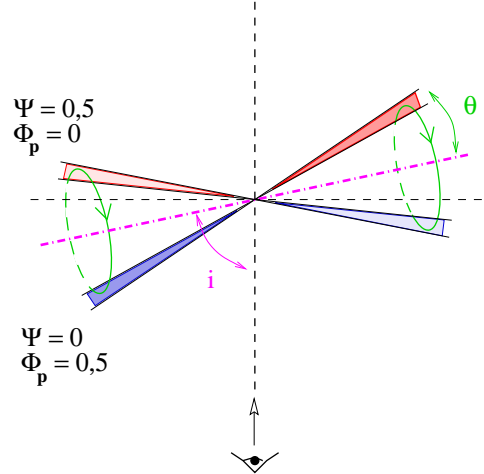}
   \caption{Convention of the precessional phase $\Phi_\mathrm{p}$
   according to \citet{kemp86} and the usual precessional phase $\Psi$
   (e.g. \citealt{goranskii98}) shown on the kinematic model of
   SS\,433: $P_\mathrm{orb}=13.08$\,d,
   $P_\mathrm{prec}=162.375\pm0.011$\,d, velocity of the ejections
   $v=0.2647\pm0.0008$\,c, inclination of the jet axis to the line of
   sight $i=78.05\degr\pm0.5\degr$ and opening angle of the precession
   cone $\theta=20.93\degr\pm0.08\degr$ \citep{eiken01}.}
    \label{figmodSS433}
    \end{figure}

        In Table~\ref{tabobsSS433}, $\Phi_\mathrm{p}$ and $\varphi$
        are, respectively, the precessional and orbital phases according
        to \citet{kemp86}. As usually $\varphi=0$ corresponds to the
        primary minimum in the visible light curve, but here the
        mid-eclipse is the donor inferior conjunction, 
	as the region of the compact object
         is more luminous than the donor star.
        \citet{kemp86} chose the precession phase $\Phi_\mathrm{p}=0$
        at the time of light minimum in terms of the 162\,d Fourier
        component, corresponding to the time when we tend to look
        ``under'' the disc, but many authors, such as \citet{goranskii98}
        and \citet{giesmcswain02} defined the precessional phase with
        $\Psi=0$ (corresponding to $\Phi_\mathrm{p}=0.5$) when the
        emission lines attain their extremum radial velocities, thus the
        time when the jets are closest to our line of sight and the
        disc presents a maximum cross section to us (see
        Fig.~\ref{figmodSS433}).
        The values obtained when calculating the orbital phase
        $\varphi$ using the ephemeris from \citet{kemp86} are quite
        similar ($\pm0.01$) to the ones obtained following the
        ephemeris of \citet{goranskii98}. There is a difference of
        $\pm0.09$ in the results obtained for the precessional phases
        using both ephemeris, probably due to the different
        precession periods used by the authors: 162.5\,d for
        \citet{kemp86}, instead of 162.15\,d for \citet{goranskii98}
        and \citet{giesmcswain02}.  

        We reported the values of the orbital and precessional phases
        of SS\,433 with Figs.~3 and~4 of \citet{kemp86}, respectively showing
        the mean V-band light curve on the orbital period and 162.5\,d
        light curves centred on four orbital phases ($\varphi=0.0,
        0.25, 0.50, 0.75$). The decreasing level with time of the
        continuum in April 1997 (see Fig.~\ref{figcont}) is easily
        explained by the decrease of the corresponding levels in the
        mean orbital light curve and in the precessional light
        curve. The continuum level of the spectrum on 3 November 1996
        is remarkably similar to the one on 17 April 1997 while the
        mean orbital light curve predicts a slightly lower level. But
        this could be equilibrated by its precessional phase which is
        a bit farther to the minimum flux than in April
        17.

        However, we cannot clearly explain why the 24 October~1997
        level is so low while its orbital phase $\varphi=0.25$
        should correspond to the maximum orbital flux, unless
        considering that the effect of the precession phase is
        dominant, since with $\Phi_\mathrm{p}=0.02\pm0.05$ this date is
        very close to the minimum flux 
        when the disc is edge on. 
        Then the much higher flux two days later, on 26 October 1997 
        should also be attributed to the slight higher precession
        phase ($\Phi_\mathrm{p}=0.03\pm0.05$), which could lead to a
        significantly higher flux because of geometrical effects. Note
        also that these observations in October 1997 were performed with a
        different instrument, ISOCAM instead of ISOPHOT, so with
        different sensitivity and spatial resolution.

   \subsection{Data fitting}
        
        The continuum of SS\,433 in April 1997 can be fitted very well
        by one or two power laws and a blackbody emission as shown in
        Fig.~\ref{figcont} (left).  For 11 April 1997 the continuum
        corresponds to a power law from 2.5 to 12\,$\mu$m: 
        $F_\nu = C\,\lambda^{-\alpha}$\,Jy, where $C=2.3$ and
        $-\alpha=-0.6\pm0.05$. Then a blackbody model with $T=150$\,K
        and $R=8000\,R_\odot$
        is added to this power law emission to
        fit the far-IR emission, but with only three points at 12, 25,
        and 60\,$\mu$m, the constraints are not strong. The blackbody
        model was the simplest model to correspond to a possible
        thermal emission in this wavelength range. The two other April
        1997 spectra are fitted by two power laws
        with $C\!=\!0.6$,\ $-\alpha\!=\!+0.1\pm0.15$ on April~17 and $C=0.35$,
        $-\alpha\!=\!+0.3^{+0.2}_{-0.15}$ on April 23 for the
        \mbox{2--4.5\,$\mu$m} range, and 
         $C=1.6$, $-\alpha=-0.6^{+0.25}_{-0.05}$ and $C=1.3$, \ 
         $-\alpha=-0.6^{+0.25}_{-0.1}$, respectively, for the
         4.5--12\,$\mu$m part.

        These power laws can be interpreted as free-free emission:
        optically thin for the positive slopes between 2 and
        4.5\,$\mu$m in April 17 and 23, and the negative slope
        ($\lambda^{-0.6}$) corresponding to the intermediate regime
        between optically thin and optically thick free-free
        emission. This $-0.6$ slope is the exact theoretical
        spectral index for the free-free emission from an ionized
        homogeneous wind flowing out from a star with a spherical
        expansion and at a constant velocity
        \citep{panagiafelli75,wrightbarlow75}. 
	This is characteristic of hot stars with strong winds, such
	as O and WR stars \citep{cohen75}.
	Note that
        \citet{schmid82} demonstrated that the overall spectral
        emitting behaviour is the same for non-spherical outflows with
        more complex geometries, as long as they stay thick, i.e. when
        no ratio of structural length scales in the source exceeds
        about 10, which is very likely the case in SS\,433.

        Figure~\ref{figcont} (right) shows our fit of the continuum of
        the 3 November 1996 spectrum with the same functions as on
        17 April 1997, as they have the same flux level: 
        $F_\nu = C\,\lambda^{-\alpha}$\,Jy with $C=0.6$, \ 
        $-\alpha=+0.1\pm0.1$ and $C=1.6$, \ $-\alpha=-0.6\pm0.15$. These
        functions fit the continuum well except for $\lambda \gtrsim
        10$\,$\mu$m, where the continuum decreases more rapidly as if
        it turns to an optically thicker emission. The same behaviour
        is observed with ISOCAM on 24 October 1997, although the wide
        band filters have large error bars. This spectral energy
        distribution follows the $F_\nu = 1.1\,\lambda^{-0.6}$\,Jy law
        between 5 and 10\,$\mu$m, then the flux decreases indicating a
        clear cut off toward a larger slope. The continuum can then be
        fitted by an optically thick free-free emission ($F_\nu =
        24\,\lambda^{-2}$\,Jy) plus the same black body emission
        ($T=150$\,K and $R=8000\,R_\odot$)
        as in the April 1997
        fitting.

        Thus over five epochs of observations, the 2--12\,$\mu$m
        spectrum of SS\,433 corresponds to optically thin and thick
        free-free emission, so it is consistent with a standard O or WR
        wind. The far-IR emission is probably thermal emission from dust
        at $T=150$\,K surrounding the system at a large distance
        \mbox{($R\gtrsim 8000\,R_\odot$).}


   \subsection{Mass loss evaluation}
        \label{sectmdot}

        From our fit of the free-free emission of SS\,433 and assuming
        a thick geometry for the wind,
        we can
        calculate the corresponding mass loss $\dot{M}$ (in $M_\odot\, 
        \mathrm{yr}^{-1}$) using the \citet{wrightbarlow75} equation (8), which
        gives the flux density $F_\nu$ of the free-free emission of
        a radiatively driven wind:
        \begin{displaymath} 
        F_\nu \, =\, \frac{2.32\times10^{10}}{D_\mathrm{kpc}^2} \ 
          \left(\frac{\dot{M}}{\mu \upsilon_\infty}\right)^{4/3} 
      \ (\nu_\mathrm{GHz}\,\gamma_\mathrm{e}\,g\,Z^2)^{2/3} \quad \mathrm{mJy}
        \end{displaymath} 
        where $D_\mathrm{kpc}$ is the distance in kpc, $\mu$  mean
        atomic weight per nucleon, $\upsilon_\infty$ the terminal velocity
        of the wind measured in km\,s$^{-1}$, $\nu_\mathrm{GHz}$ the
        frequency of observation in GHz, $\gamma_\mathrm{e}$ the
        number of free electrons per nucleon, $g$ the Gaunt factor
        and $Z$ the mean ionic charge. Then the mass loss is
        \begin{displaymath} 
        \dot{M}\, =\, 1,69.10^{-8} \, F_\nu^{3/4}\,D_\mathrm{kpc}^{3/2} \ 
         \mu \upsilon_\infty \, 
        (\nu_\mathrm{GHz}\, \gamma_\mathrm{e}\,g\, Z^2)^{-1/2}\,
                \quad M_\odot \ \mathrm{yr}^{-1} .
        \end{displaymath}
         We took $D=5.5$\,kpc
        and we measured the flux densities at a
         wavelength of 5\,$\mu$m ($\nu=60\,000$\,GHz) in the maximum
         and minimum levels of April 1997: $F_\nu= 876$\,mJy \& 495\,mJy.
         The other parameters of this equation are not very well
         known, but we can take either the usual or average values of the WN
         winds, assuming that the Gaunt factor is $g=1$ 
        and that the terminal
         velocity is $\upsilon_\infty=1000$\,km\,s$^{-1}$ \citep{crowther03},
         and using a wind composition typical of late WN stars
         $\mu=2$, $\gamma_\mathrm{e}=1$, and $Z=1$
         \citep{leitherer97}.
	Note that the adopted velocity is also compatible with the 
	observations of the H$_\alpha$ stationary lines in the optical 
	spectrum; \citet{vandenheuvel81} reported Doppler widths of 
	$\sim 1-2 \times 10^3$~km\,s$^{-1}$, and recently \citet{giesmcswain02}
	measured FWHM of $\sim$\,15\,\AA\ ($\sim$\,680\,km\,s$^{-1}$).
        We find a maximum mass loss rate of
        $\dot{M}=2.9\times 10^{-4} \ M_\odot \,\mathrm{yr}^{-1}$
        and a minimum of 
        $\dot{M}=1.9\times 10^{-4} \ M_\odot \,\mathrm{yr}^{-1}$.

        We explored the possible range of the parameters to see how
        they influence the resulting mass loss. The velocity of the
        wind flowing out from a WR star generally lies between
        500\,km\,s$^{-1}$ and 1500\,km\,s$^{-1}$, which either reduces
        the result by a factor of 2 or multiplies it by 1.5. 
        In the mid-IR, the Gaunt factor is not easy to calculate, as
        the parameters of the plasma are unknown. According to
        \citet{rybickilightman79}, $g$ ranges from 1 to 5, which
        would result in a mass loss multiplied by 0.45.
        Another unknown is the chemical composition of the wind, but
        the typical WN composition seems reasonable since the study of
        \citet{leitherer95} shows values of $\mu$ between 1.5 and 2.7
        for WN7/WN8 stars (which implies $\dot{M} \times0.75$ to 1.35);
        and from the mid-IR spectra, the wind is dominated by H$^+$ and
        He$^+$ ions so $\gamma_\mathrm{e}=Z=1$.
        We checked the results obtained by dereddening the spectra
        with a lower visible absorption $A_V=5.3$ and found that this
        does not change the mass loss a lot, and it is then multiplied by a
        factor of 0.85.
        We also calculated the mass loss using 
        the lowest value of distance $D=3.2$\,kpc,
        inferred from \ion{H}{i} observations of the W50 nebula around
        SS\,433 \citep{dubner98}, which divides the resulting
        mass loss values by a factor of 2.2.
%
        Thus the greatest uncertainties concerning the mass loss come
        from the wind velocity and the Gaunt factor, so that the range
        $2-3 \times 10^{-4} \ M_\odot \,\mathrm{yr}^{-1}$ 
        is valid within a factor of 3. 

        This result agrees well with the past
        estimates. \citet{vandenheuvel81} found a wind mass loss rate
        of $10^{-5}-10^{-4} \ M_\odot \,\mathrm{yr}^{-1}$ outflowing
        from an O or WR type star with a velocity of
        $1000-2000$\,km\,s$^{-1}$. \citet{shklovskii81} calculated a
        mass loss rate of $\sim 10^{-4} \ M_\odot \,\mathrm{yr}^{-1}$
        from a 10\,$\mu$m flux density and by assuming an optically thick
        plasma with a dispersion velocity of $\upsilon \sim
        2000$\,km\,s$^{-1}$. More recently, \citet{king00} considered
        possible evolutionary models for SS\,433 and predicted mass
        loss rates between $7 \times 10^{-6} \ M_\odot
        \,\mathrm{yr}^{-1}$ and $4 \times 10^{-4} \ M_\odot
        \,\mathrm{yr}^{-1}$.

        We can also compare this result with our previous study of
        Cygnus\,X-3 \citep{koch02}, which is 
        generally accepted as an X-ray
        binary containing a Wolf-Rayet star. 
        The only other such case is IC\,10\,X-1 \citep{clark04}.
        In this similar study,
        the continuum of the mid-IR spectrum of Cygnus\,X-3 is also
        due to free-free emission and corresponds to a mass loss rate
        of $1.2 \times 10^{-4} \ M_\odot\,\mathrm{yr}^{-1}$ when using
        the same typical WN composition and Gaunt factor and with a
        velocity of $\upsilon_\infty=1500$\,km\,s$^{-1}$. The similarity of
        both results is striking, whereas the observed flux is fairly
        different, since \mbox{Cygnus\,X-3} is at a distance of 10\,kpc with a
        highly absorbed emission ($A_V=20$). 
        This shows that the spectrum of SS\,433 resembles
	the one of a
        Wolf-Rayet X-ray binary very well
        and suggests a common
        physical mechanism at work in both systems.
        Indeed a disc-like wind, 
        though significantly flattened in the plane of the
        binary orbit, has been invoked by \citet{fender99} to explain both 
        the variability of the \ion{He}{i} emission lines in this
        object and an asymmetric emitting region. 
        This binary
        system has a much smaller size ($\sim5\,R_\odot$ separation,
        4.8\,h period) than SS\,433 (13\,d period), thus the IR emitting
        region is significantly larger than the binary separation, 
        consequently the outflowing wind is external to the binary
        orbit. 


        Finally, the mass loss rate of SS\,433 has to be corrected
        from the effect due to the very probable inhomogeneity of the
        wind. The clump mass loss rates of WN stars are a factor of 3 times
        lower than previous homogeneous rates \citep{crowther03}, so
        the mass loss rate of SS\,433 is within the range
        $6-10\times 10^{-5} \ M_\odot \,\mathrm{yr}^{-1}$.
        This result can be compared to the mass loss rate 
	$1.5-3.7 \times 10^{-5} M_\odot \,\mathrm{yr}^{-1}$ of WR\,147
        measured by \citet{morris00} and to the typical range of WN clumped
        mass-loss rates: $10^{-5.5}-10^{-4.5}=3.16 \times 10^{-6}-3.16
        \times 10^{-5} M_\odot \,\mathrm{yr}^{-1}$ \citep{crowther02}.
        Thus the mass loss rate
        found for SS\,433 is 
        the same order of magnitude as a strong WN wind.


   \begin{figure*}[!t]
        \includegraphics[width=6cm]{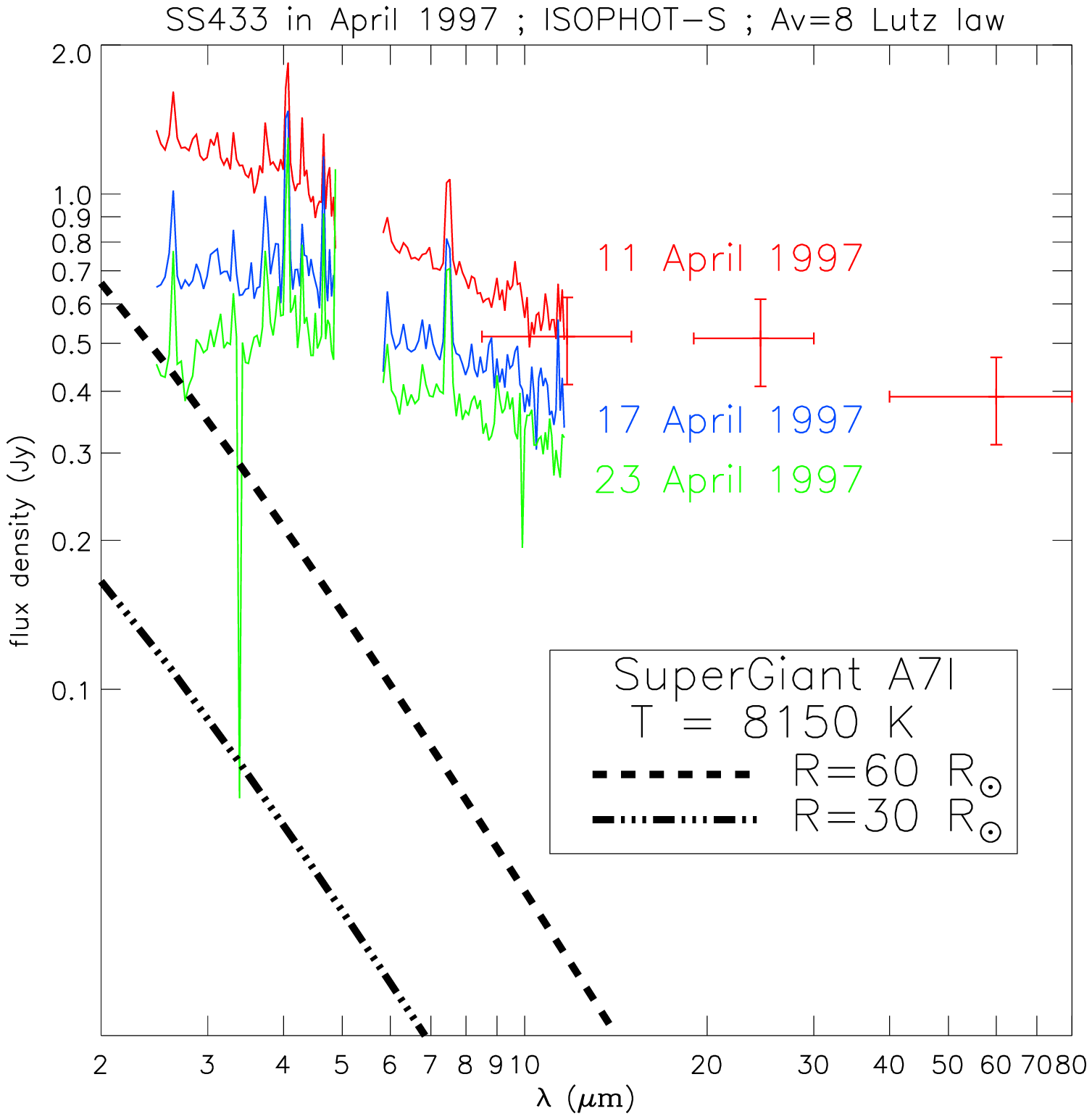}
        \includegraphics[width=6cm]{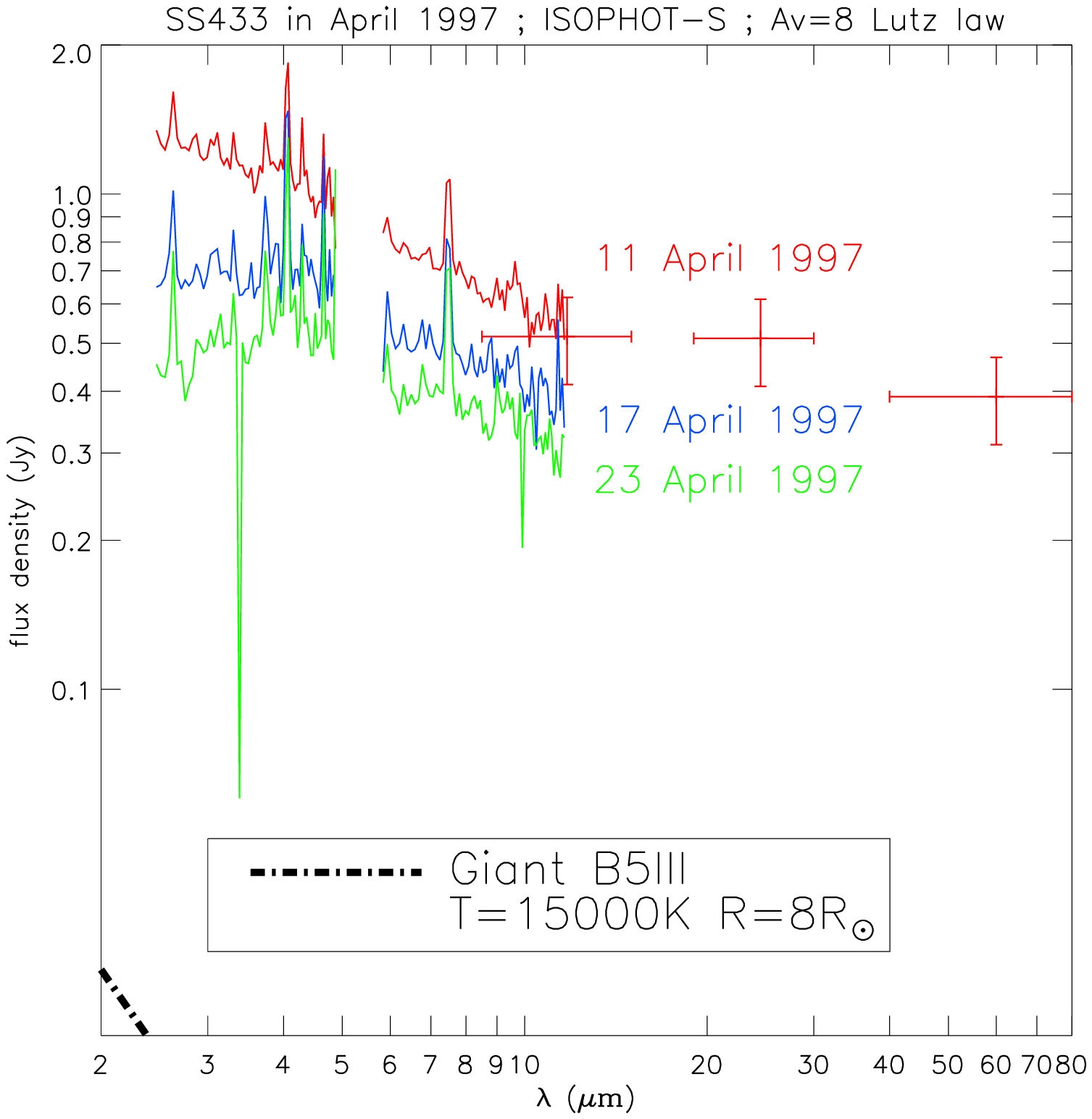}
        \includegraphics[width=6cm]{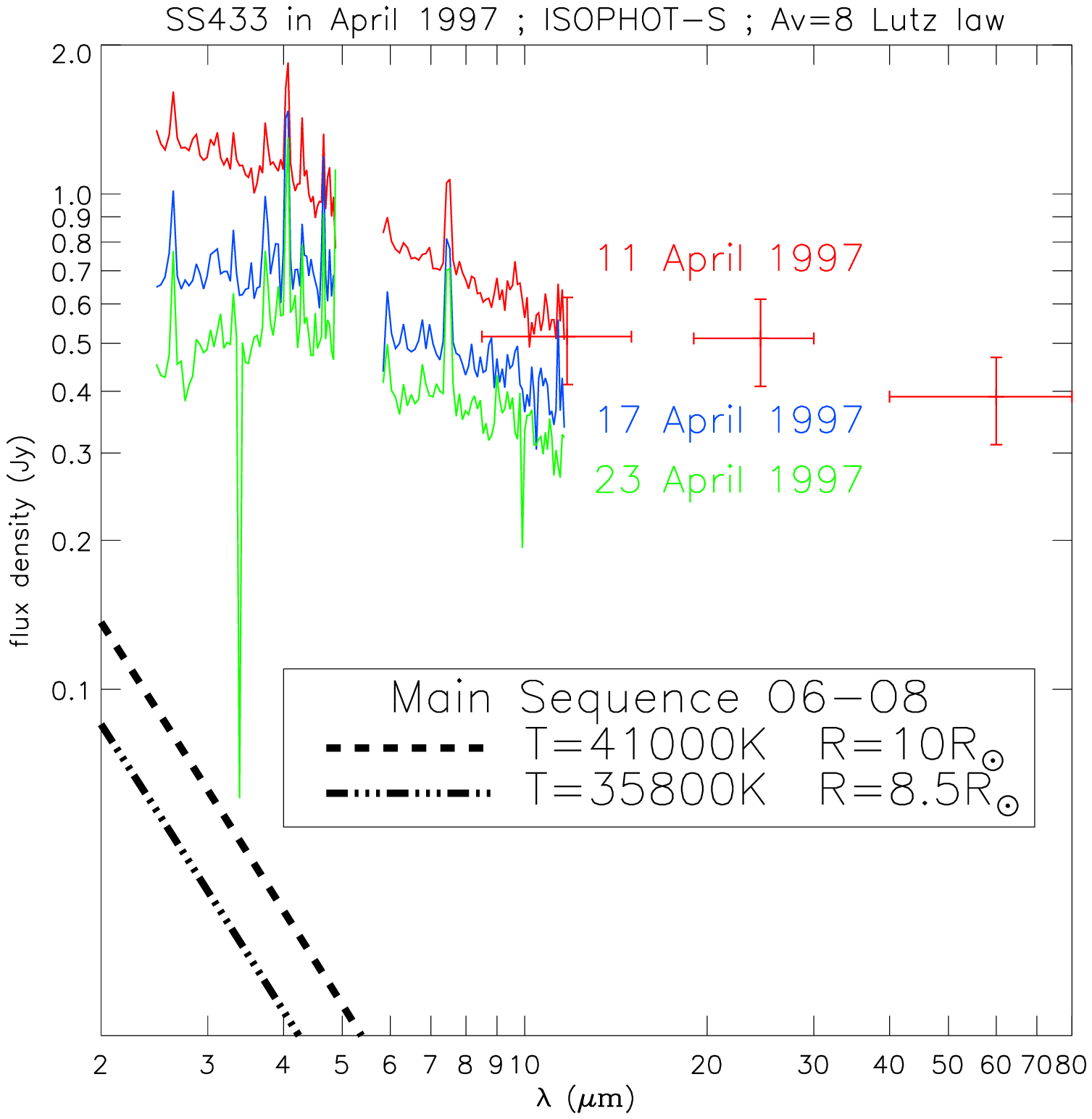}
   \caption{Superimposition of the ISOPHOT mid-IR emissions of
   SS\,433 with the flux from the possible donor stars modeled with
   black body emissions at 5.5\,kpc: 
   on the left side an A-type supergiant with $T=8150$\,K and
   $R=60$ (dashes) or $30\,R_\odot$ (dash triple dot), 
   at the centre a B5 giant with $T=15000$\,K and $R=8\,R_\odot$ (dash dot) 
   and on the right side a main sequence O6 or O9 star with
   $T=41000$\,K and $R=10\,R_\odot$ (dashes) or $T=41000$\,K and 
   $R=8.5\,R_\odot$ (dash triple dot), respectively.}
   \label{figconstraint}
   \end{figure*}

   \subsection{Size of the emitting region}
        \label{sectrayon}

        Although we don't know the geometry of this wind, we can
        calculate the radius of the equivalent spherical free-free
        emitting region thanks to Eq. (11) of
        \citet{wrightbarlow75}:
        \begin{displaymath} 
        R(\nu) \, =\, 2.8\times10^{28} \, \gamma_\mathrm{e}^{1/3} \, g^{1/3}
          \, Z^{2/3} \, T_\mathrm{e}^{-1/2} \, 
          \left(\frac{\dot{M}}{\mu \upsilon_\infty \nu}\right)^{2/3}
          \quad \mathrm{cm}
        \end{displaymath} 
        where $T_\mathrm{e}$ is the electron temperature of the gas
        (in Kelvin), which is assumed to be constant.  Combining this
        equation with Eq.~(8) of \citet{wrightbarlow75} (see
        Sect.~\ref{sectmdot}) gives the characteristic radius of the
        emitting region at a frequency $\nu$:
        \begin{displaymath} 
        R(\nu) \, =\,1.84\times10^{17}\,T_\mathrm{e}^{-1/2}\,D_\mathrm{kpc}\, 
        \nu_\mathrm{GHz}^{-1} \, F_{\nu,\,\mathrm{mJy}}^{1/2} \quad \mathrm{cm}.
        \end{displaymath} 
        The temperature of the plasma responsible for the observed
        free-free emission is unknown, so we take the typical value of
        $T_\mathrm{e}=10\,000$\,K for WR stars according to
        \citet{leitherer95}. At 5\,$\mu$m $=6\times 10^4$\,GHz with
        the minimum (495\,mJy) and maximum (876\,mJy) flux densities
        of April 1997, we find 
   	$R_{5\mu\mathrm{m}}=3.8-5.0 \times 10^{12}$\,cm $=50-67\ R_\odot$.

        The value of this characteristic radius is very dependent on
        the frequency of observation, but also on the plasma
        temperature and on the distance. If the plasma temperature
        varies by a factor of 2 or 3, this radius will be multiplied
        by a factor of 0.7 or 0.6, respectively. The lower distance
        3.2\,kpc will multiply this result by a factor of 
        0.6, and the
        lower visible extinction will only multiply it by a factor of
        0.9. Moreover, this is the equivalent radius referring to a
        spherical case, so these values can only be taken into account
        as orders of magnitude for the typical size of the free-free
        emitting region in SS\,433.

        This result can be compared to the previous study of
        \citet{shklovskii81}, who used a measurement at 10\,$\mu$m to
        find that the linear radius of the emitting region (considered as
        optically thick) is $R=5\times 10^{12}$\,cm (72\,$R_\odot$),
        which is compatible with our result. He corrected this with
        the effects of inclination of the system to the line of sight,
        considering the emitting region as an oblate spheroid whose
        principal radius is thus $10^{13}$\,cm $\simeq 144\,R_\odot$,
        thus greater than the binary system and forming a common
        envelope around it. These kinds of models will be discussed later.


\section{Discussion}

   \subsection{Nature of the donor star and mid-IR constraints}

        In the previous section we showed that both the emission lines
        and the continuum of the mid-IR spectrum of SS\,433 are
        compatible with a Wolf-Rayet type for the donor star and
        preferably with a late WN star (WN8).
%
        \emph{However}, the recent works of \citet{gieshuang02} and
        \citet{hillwig04} suggest that the donor star is an A3-7\,I
        supergiant star with $10.9 \pm 3.1 \, M_\odot$. This result
        seems to be confirmed by the preliminary study of blue spectra
        of SS\,433 by \citet{charles04}, who find absorption features
        typically seen in late B-A spectra. On the other hand,
        \citet{lopezmarshall04} use a different method, modeling the
        Chandra X-ray spectrum during eclipse to constrain the radius
        of the mass donor star to be $9.1 \pm 1.0 \, R_\odot$. They
        note that for a main sequence star, this corresponds to a
        companion mass of $29 \pm 7 \, M_\odot$, so an O6-O8\,V type
        star with $T=41000-35800$\,K, according to \citet{lang92}. But
        such a radius could also correspond to a \mbox{B4-5} giant star. For
        a B5\,III ($R=8\,R_\odot$), we find $M=7\,M_\odot$ and
        $T=15000$\,K \citep{lang92}. 
        To sum up, following all these results
         the donor star is not
        a Wolf-Rayet star, so there must be a phenomenon imitating such a
        star in SS\,433.

        Before we discuss this point, let us see which constraints on
        the nature of the donor star we can provide from the mid-IR
        spectra of SS\,433. In Fig.~\ref{figconstraint} we overplotted
         the emissions of the possible
        donor stars above cited to the dereddened ISO spectra. 
	We modeled these stars with black
        body emissions at a distance of 5.5\,kpc.

        Concerning the A\,I
        hypothesis, we chose to fit an A7 supergiant which is less
        luminous ($T=8150$\,K from \citealt{lang92}) than an A3\,I
        ($T=8770$\,K), and we took a stellar radius of $R=60\,R_\odot$
        \citep{lang92}.
        This emission is not compatible with the
        mid-IR emission of SS\,433, since in the $2-5\,\mu$m range the
        A7\,I flux density is 
	of the order of the continuum
        of SS\,433 on 23 April 1997 (Fig.~\ref{figconstraint}
        left). But this continuum corresponds to an optically thin
        free-free emission, which should then not be detected because
        of the dominant black body emission at this orbital phase
        ($\varphi = 0.16$) where the donor star is in the front part
        of the orbit. In order to get a black body emission negligible
        compared to this optically thin emission, we had to set the
        radius of the A7\,I star to $R=30\,R_\odot$ or less. Then this
        constraint is compatible with the result of \citet{hillwig04},
        who find the radius of the Roche lobe volume for the mass
        donor star to be $R_L = 28 \pm 2 \, R_\odot$. Note that this
        corresponds to a star with an intermediate size between type I
        (supergiant) and type II (bright giant) stars \citep{venn95}.


        Concerning a possible donor star with $R=9.1 \pm 1.0 \,
        R_\odot$, we used the standard values for the radius $R$ and
        temperature $T$ of a B5\,III and O6-O8\,V stars, and
        Fig.~\ref{figconstraint} (centre and right) shows that the
        corresponding black body emissions are negligible compared to
        the flux density of the continuum obtained with ISOPHOT. Thus
        both a giant B5 star and a main sequence O6-O8 star are
        compatible with the spectrum of SS\,433.

   \subsection{A phenomenon imitating a Wolf-Rayet star}

        Now, as
        the mass donor star of SS\,433 is not a Wolf-Rayet
        star, then there must be a phenomenon imitating the emission of
        such a star in this binary system. 
%
        A long time ago, \citet{vandenheuvel81} suggested that
        SS\,433 was a
        system containing an early-type star overflowing its
        Roche-Lobe and a large luminous accretion disc, both
        embedded in a dense, spherically
        symmetric, outflowing atmosphere resembling that of a
        Wolf-Rayet star.
%
        We propose a very similar model 
        where this Wolf-Rayet
        phenomenon comes from the material surrounding the compact
        object and where the donor star is only constrained to be 
	smaller than a supergiant. 
	This material does not form a classical thin accretion
        disc but instead a thick torus or a thick envelope like a stellar
        atmosphere, but not spherical,
        which is ionized by the UV/X-rays emitted in the close
        vicinity of the compact object and expelled
        by radiation pressure, thus imitating the wind of a Wolf-Rayet
        star.

        There may be a ``classical'' thin accretion disc, but only
        very close to the compact object and not detectable because of
        the surrounding material.
        This would explain that there is no evidence for an accretion
        disc in the Chandra X-ray spectrum of SS\,433 according to
        \citet{marshall02} and that SS\,433 is not as strong an X-ray
        source as the other microquasars or X-ray transients.


        Note that \citet{hillwig04} find a binary separation of
        $a=0.26 \pm 0.02$\,AU $=56 \pm 4\,R_\odot$, which is of the
        order of the typical radius of the free-free emitting region
        at 5\,$\mu$m 
        $R_{5\mu\mathrm{m}}=50-67\ R_\odot$, which we
        calculated in Sect.~\ref{sectrayon}. So the wind may envelop a
        part of the donor star or the whole binary system, possibly
        forming a very hot ``common envelope'', as already suggested
        by \citet{shklovskii81}. 
%

        Several authors have reported the possibility of a peculiar
        envelope surrounding SS\,433 with a disc-like shape.
        \citet{zwitter91} proposed that the asymmetric shape of the
        visible light curves is due to a disc-like outflow of matter
        from the accretion disc. This optically thick ``excretion
        flow'' is not confined to the plane of the accretion disc but
        has an opening angle of $\sim 20\degr$ centred close to the
        orbital plane and extends up to the orbit of the companion.
        This analysis is compatible with the observations of
        \citet{kotani96}, who explained that some X-ray lines are
        heavily absorbed only in the receding jet because of the
        existence of absorbing gas in a precessing plane perpendicular
        to the jets. However, they do not constrain the size of this
        extended rim of the accretion disc, and their evaluation of the
        mass escape rate (a few $10^{-8}-10^{-7}\
        M_\odot\,\mathrm{yr}^{-1}$) is far too low compared to our
        result based on a spherical model \citep{wrightbarlow75},
         which is also valid for a thick disc-like outflow 
        according to \citet{schmid82}. 
        Moreover, our observations suggest a thicker envelope
        with roughly the size of the SS\,433 binary system, more
        torus-like than disc-like.
        


   \subsection{Ejected material at larger scale}

        With such a huge mass loss for SS\,433 
        ($\sim 8\times 10^{-5} \ M_\odot\,\mathrm{yr}^{-1}$), 
        one could ask what happens to
        all this ejected material. The relativistic jets of SS\,433
        are well known, and the source is continuously ejecting material
        in this way, but \citet{marshall02} and previous studies
        estimated the mass loss of these jets to 
	$\sim10^{-7}\,M_\odot$\,yr$^{-1}$. 
	Thus the material ejected in the jets
        is negligible compared to the wind that we observe in the
        infrared.
        
        A part of this wind material might form dust, which 
	emission in the far infrared would explain the
	excess emission that we observe
        for $\lambda > 15\,\mu$m in the 11 April 1997
        spectrum of SS\,433 
        (Fig.~\ref{figcont}). We modeled this emission with a black
        body of temperature $T=150$\,K and radius 
        $R=8000\,R_\odot=5.6\times 10^{14}$\,cm $= 37.2$\,AU,
        which gives a rough
        estimate of the temperature and distance of the dust region.
        No WN star is known to form dust, so this would be
        another surprising difference between SS\,433 and
	WN stars. This dust
        could form when the ejected material cools down, far enough
        from the binary system so that it is not destroyed by the hot
        radiation of the WR-like object. SS 433 would then resemble  
	some planetary nebulae
        where the central star can be very hot ($10^4-10^5$~K), 
	particularly the newly discovered [WN7] 
	star associated with the ring nebula PCG11 \citep{cohen05}. 
	The visible spectrum of SS\,433 clearly shows C lines:
	\ion{C}{iii}/\ion{N}{iii} 
	$\lambda$4644\,\AA\ \citep{margon84,gieshuang02}
	and \ion{C}{ii} $\lambda\lambda$7231,7236\,\AA\ \citep{giesmcswain02}.
	The
        abundance of C may not be high in the wind of SS\,433, but the
        mass loss is so huge that it brings a lot of material to the
        surrounding medium, so the conditions for dust formation might
        be gathered in the outskirts of SS\,433. 

        \citet{fabrika93} argues that the gas flows out through the
        outer Lagrangian point L$_2$ behind the compact star and
        sprinkles a spiral of matter forming a disc-like envelope with
        an opening angle of about $20\degr$. In the course of
        precession, an extended disc is formed with a double-cone
        shape with an opening angle of$\sim60\degr$. Its inner radius
        is $10^{14}$\,cm, and it is not fully filled by the gas up to a
        distance of $10^{15}$\,cm. According to this model, in
        $10^{4}$\,years the external radius of this envelope is
        1.5\,pc.
        Depending on the distance of the L$_2$ point to the compact
        object (so depending on its mass) this analysis is compatible
        (or not), with our thick wind envelope.

        This scenario is also -- in general -- compatible with the
        radio observations of equatorial outflows in SS\,433. In the
        1.6\,GHz image of \citet{paragi99}, there are two faint
        extended emission regions at a distance of 30\,mas (138\,AU) to
        70\,mas (322\,AU) to the NE and SW from the position of
        SS\,433. This detection is confirmed by \citet{blundell01} who
        show smooth low-surface brightness emissions extending to
        $\gtrsim 40$\,mas perpendicularly to the jet at 1.5\,GHz and
        5\,GHz. More observations analyzed by \citet{paragi02} show
        that these radio features
        extending along the equatorial plane of the binary system move
        away from the central engine on timescales of weeks to
        months. It seems that the appearance of these radio emitting
        regions is changing with time, blob-like or smoother, and is
        related to the precessional cycle. The nature of the emission
        is unknown, very likely non-thermal and maybe optically thin
        synchrotron from a mixed population of relativistic and
        thermal electrons.

        This equatorial outflow of SS\,433 has also recently been
        observed with the VLBA by \citet{mioduszewski04}. The diffuse
        emission of the outflow seems to brighten as it moves away
        from SS\,433 at a velocity of $\sim 5000-10000$\,km\,s$^{-1}$.
        This is much higher than the wind velocity that we 
	adopted in
        our study and even much higher than the upper limit for the
        velocity of a WR wind. Thus, either the wind outflowing from
        SS\,433 has a very unusual high velocity -- which we cannot
        explain and which would increase the mass loss rate by nearly
        an order of magnitude -- or the wind material 
        (mainly electrons) is accelerated
        with an unknown (magneto-hydrodynamic?) mechanism somewhere
        between the binary system and the large distance ($>100$\,AU)
        of the radio emitting regions.



\section{Conclusions}

        The mid-IR spectra of SS\,433 obtained with ISO in 1996 and
        1997 show H and He emission lines similar to the ones observed
        in the spectrum of the Wolf-Rayet star WR\,147, a WN8(h)+B0.5V
        binary system with colliding wind. The spectrum of SS\,433 is
        thus compatible with the presence of a WN8 
        star.

        The 2--12\,$\mu$m continuum emission of SS\,433 corresponds to
        free-free emission, optically thin or intermediate between
        optically thin and thick, depending on both the wavelength and time
        of observation. At 25\,$\mu$m and 60\,$\mu$m, the emission 
        may be due to dust at
        $T\sim150K$ surrounding the binary system at a large
        distance ($\gtrsim 8000\,R_\odot$).
%
        Assuming that the free-free emission is emitted by a
        geometrically thick homogeneous wind, we calculated the
        corresponding mass loss 
        of 
        $\sim 2-3 \times 10^{-4} \ M_\odot \,\mathrm{yr}^{-1}$. 
        If the wind is
        clumped, this result is a factor of 3 times lower: 
        $\sim 6-10 \times 10^{-5} \ M_\odot \,\mathrm{yr}^{-1}$, 
        thus compatible
        with a strong wind from a WN star.

        However, considering recent results discarding a WR star for
        the nature of the mass donor star in SS\,433, we propose that
        the WR-like wind observed in the IR is outflowing from an
        envelope of material enshrouding the region of the compact
        object and maybe also the donor star. This envelope, heated,
        ionized, and expelled by the X-ray emission of the compact
        object, is thus imitating a Wolf-Rayet star.
        This wind is probably the source that provides material which
        forms the possible dust emitting in the far-IR and, at larger
        distance ($>100$\,AU), the equatorial outflow observed in radio.\\



\begin{acknowledgements}
        The ISOPHOT spectra presented in this paper belonged to the
        observing programme of Alberto Salama, and the authors thank
        him for investing quite a lot of efforts in preparing
        the observations.
%
        The authors thank Phil Charles, Simon Clark, 
        and Martin Haas for interesting 
        discussions and suggestions.
        Y.F. was partly supported by a CNES external post-doctoral fellowship.
        The work was partly supported by the grant OTKA T\,037508 of the
     Hungarian Scientific Research Fund. P.\'A. acknowledges the support
     of the Bolyai Fellowship.
        The observations were reduced using the ISOPHOT Interactive Analysis
     package PIA, which is a joint development by the ESA Astrophysics
     Division and the ISOPHOT Consortium, lead by the Max-Planck-Institut
     f\"ur Astronomie (MPIA).
        The ISOCAM data presented in this paper were analysed using
        ``CIA", a joint development by the ESA Astrophysics Division
        and the ISOCAM Consortium. The ISOCAM Consortium is led by the
        ISOCAM PI, C. Cesarsky.
        The Green Bank Interferometer is
        a facility of the National Science Foundation operated by the
        National Radio Astronomy Observatory, in support of USNO and
        NRL geodetic and astronomy programs, and of NASA High Energy
        Astrophysics programs.  
\end{acknowledgements}

\bibliographystyle{aa}
\bibliography{refss433}

\end{document}